# Multidisciplinary learning at the University scientific museums: the Bunsen burner

Aurelio Agliolo Gallitto, Vitalba Pace, Roberto Zingales

*Department of Physics and Chemistry, University of Palermo, Italy*

**Abstract**

We report on a laboratory activity carried out together with secondary school (high-school) students, with the aim of increasing their interest toward historical scientific instruments and stimulate their approach to scientific knowledge. To this purpose, we propose a hands-on activity that can be fruitfully performed at the University scientific museums. We organized a one-week summer stage at the Historical Collection of Physics Instruments and at the Museum of Chemistry of the University of Palermo. A group of selected students attended it, under the tutoring of university researchers. They were showed some Bunsen burners belonging to the collections, how they are restored, how they work and how they were used in the chemical laboratories. At the end of the stage, students introduced museum visitors to these instruments, describing them and referring about the activities they had carried out.

**Key words:** history of science, Bunsen burner, secondary school

**Corresponding author**

Prof. Aurelio Agliolo Gallitto
Department of Physics and Chemistry, University of Palermo
via Archirafi 36, I-90123 Palermo, Italy
Tel: +39 091 238.91702  Fax: +39 091 6162461
E-mail: aurelio.agliologallitto@unipa.it

# 1. Introduction

In order to increase the interest of secondary-school students in scientific degree courses, in Physics, Chemistry and Mathematics, several actions have been put forward (Fiordilino & Agliolo Gallitto, 2010). Hands-on science in research laboratories is an example (Agliolo Gallitto, 2009; Agliolo Gallitto et al., 2011). In order to provide additional opportunities that will inspire students to explore careers in scientific fields, we propose a hands-on activity that can be fruitfully carried out at the University scientific museums.

The University of Palermo has several scientific museums and historical collections. Among them, the Historical Collection of the Physics Instruments and the Museum of Chemistry are displayed at the Department of Physics and Chemistry. The oldest instruments of the collections date back to the early XIX century, when experimental Physics and Chemistry began to be taught in University courses, by using instruments and apparatus, during lectures.

Since 2014, the Department of Physics and Chemistry has organized a one-week summer Stage for secondary-school students. The aim is to increase the interest of young students toward scientific studies by utilizing laboratory activities connected with historical instruments at University scientific museums (Aglione, 2013; Barbacci, 2012). Selected students this year from the Classical Lyceum "Umberto I" and the Institute of Second-Level Education I.I.S.S. "Damiani Almeyda - Crispi" of Palermo, between the ages of 16 and 18, attended the Stage, focusing the activities on the study, conservation and valorization of instruments of historical and educational interest. The activity proposed has been devoted to Bunsen burner; it was the 'flywheel' for discussing the basic physical principle and its application in the investigation of chemical elements by flame tests. At the end of the Stage, students illustrated the instruments and the work they have done during the Stage to museum visitors.

In this article, we report on the activity carried out and discuss its pedagogical aspects. Further, we discuss the possible educational benefits toward the curricular topics of Physics and Chemistry. Our aim is to show that hands-on activity can be fruitfully carried out at the University scientific museums. We also describe some Bunsen burner models belonging to the Historical Collection of the Physics Instruments and to the Museum of Chemistry at the University of Palermo.

## 2 The origin and the development of the Bunsen burner

The gas burner, as laboratory tool, was first proposed by Michael Faraday (1791 - 1867) (Jensen, 2005; Kohn, 1950). Later, in 1857, the German chemist Robert Wilhelm Bunsen (1811 - 1899) described in detail the burner that today bears his name. It was properly designed to obtain a color-less and soot-free flame, of constant size, to be used to establish photometric standards (Jensen, 2005; Kohn, 1950). The Bunsen burner is based on the so called Venturi effect (Venturi, 1797), after the Italian physicist Giovanni Battista Venturi (1746 - 1822).

### *2.1 Venturi effect*

In 1797, Venturi reported that a fluid flowing through a constriction in a pipe undergoes a pressure reduction and an increase in speed; as a consequence, it can drag fluid along with it from a second pipe (Venturi, 1797). More generally, in a pipe of changing size, the pressure at various points is the lowest where the fluid speed and pipe constriction are the greatest, as shown in figure 1. Venturi effect can be demonstrated by pressure and velocity equations derived from the Bernoulli principle.

To show Venturi effect, the following experiment can be done. Blow into a plastic bag about one meter long and 20 cm large. If one blows with the mouth attached to the bag's open side, the amount of air that goes in will approximately correspond to the amount contained in our lung (about 4 liters). Conversely, if one blows into the bag about 10 cm far from its open side, surprisingly much more air will go inside the bag inflating it immediately.

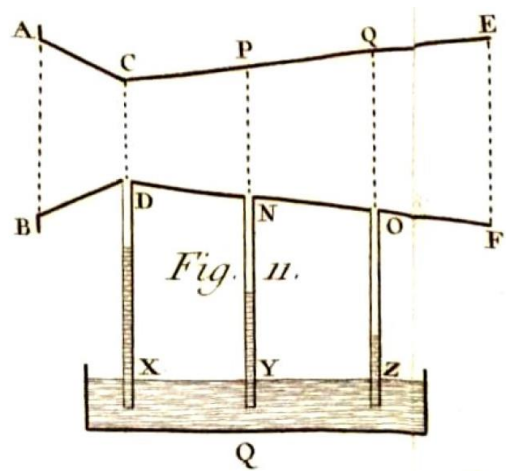

**Fig. 1.** Pressure at various points in a pipe of changing size, indicated by the height of the liquid in the vertical straws, from Venturi, 1797.

## 2.2 Hydraulic vacuum pump

The reduction in fluid pressure due to Venturi effect is the foundation of several instruments. An application is the hydraulic vacuum pump; its scheme, illustrated in figure 2a, shows the constricted section of the pipe and the low-pressure chamber. In figure 2b it is shown the hydraulic vacuum pump, made of chrome-plated brass, of the Historical Collection of Physics Instruments. It was purchased in 1936 by Emilio Segrè (1905 - 1989), during his stay in Palermo as Director of the Istituto di Fisica (today Department of Physics and Chemistry) of the University.

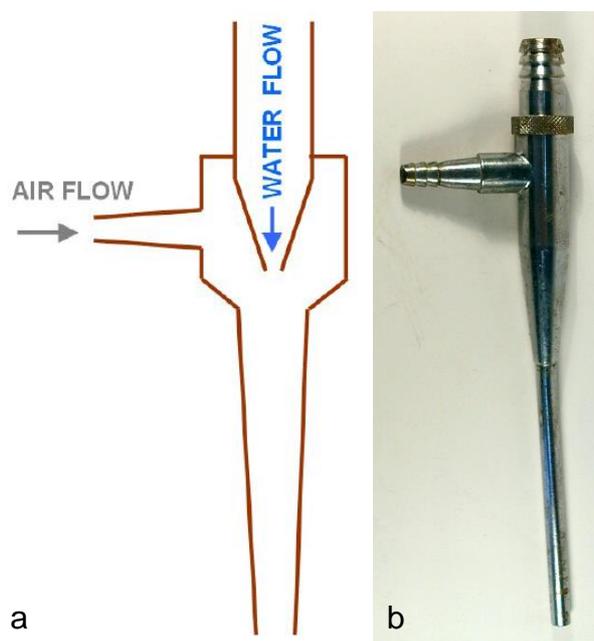

a                b

**Fig. 2.** a) Scheme of the hydraulic vacuum pump. b) Hydraulic vacuum pump made of chrome-plated brass (5 x 2 x 35 cm), purchased by Emilio Segrè in 1936 (Inv. N. 929).

## 2.3 Bunsen burner

The Bunsen burner is also based on the Venturi effect; its scheme is shown in fig. 3. When gas flows through the nozzle, its speed increases, while its pressure decreases, giving rise to an inward air flow, through the lateral holes at the base of the vertical pipe (barrel). The air inlet is adjusted by rotating a sliding collar, thus opening or closing the holes at the base, whereas gas flow is regulated by the tap in the pipe for gas inlet, at the base of the burner. If methane is used as fuel, the following combustion reaction takes place, in which $CO_2$ gas and $H_2O$ vapor are produced

$$CH_4 + 2\,O_2 \rightarrow CO_2 + 2\,H_2O$$

The structure of the flame produced by this kind of burner was described by Bunsen himself. It consists essentially of two parts, as shown in figure 3. An inner blue cone with a luminous blue tip consisting of gas which has been only partially burnt, due to a poorer air content in the gas mixture. An outer non-luminous, almost invisible, flame, consisting of completely burnt gas, due to the optimal air content in the gas mixture. For chemical analysis, a dry sample is put in the hottest zone of the flame that will get the characteristic colour of the chemical element under investigation.

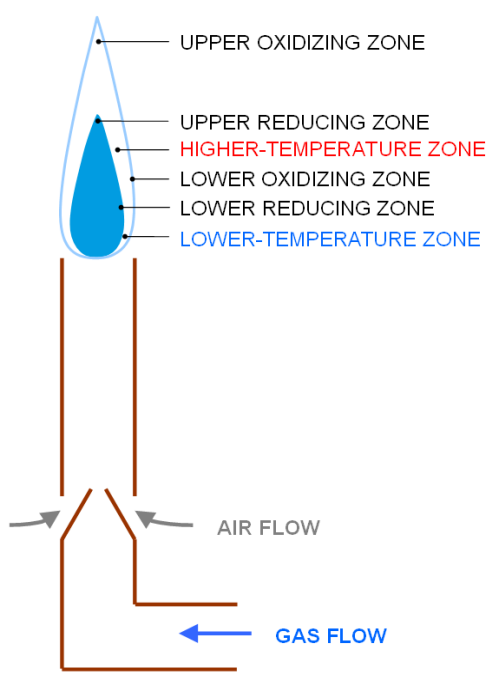

**Fig. 3.** Scheme of a Bunsen burner and its non-luminous flame; the hottest zone of the flame can reach a temperature of about 1500°C.

## *2.4 The burners of the collections*

The Historical Collection of the Physics Instruments owns several gas burners. In the inventory registry, it is reported the purchase of N. 4 gas lamp in 1887 (Inv. N. 413), N. 6 gas burners in 1889 (Inv. N. 435) and N. 1 Bunsen gas burner in 1904 (Inv. N. 549). Nevertheless, the inventory number is no longer reported on the instruments. Figure 4a shows the simple Bunsen burner. Figure 4b shows the Bunsen burner with needle valve, for fine gas flow adjustment. Figure 4c and 4d shows the burner with a lever tap, for gas flow regulation, and an inner thin straw, for an auxiliary flame. The latter can be dated at the beginning of XX century, indeed a very similar Bunsen burner is described in the Lautenschlager's Catalog of 1910 ca.

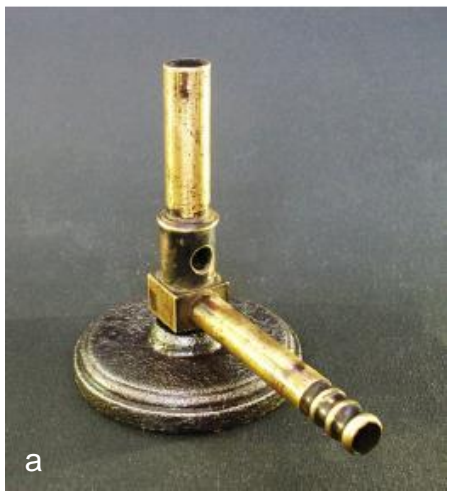
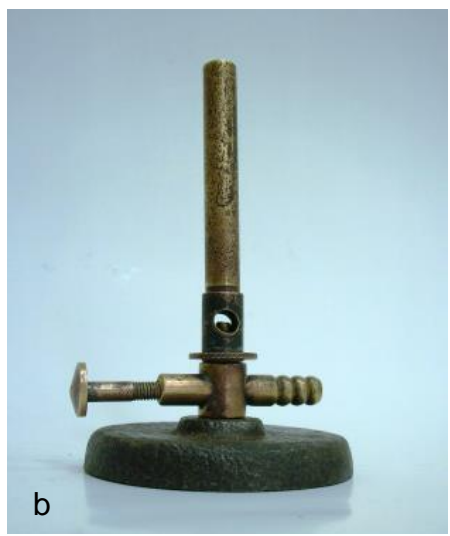
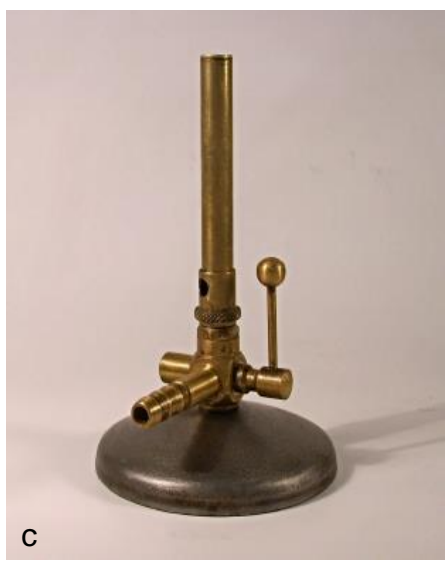
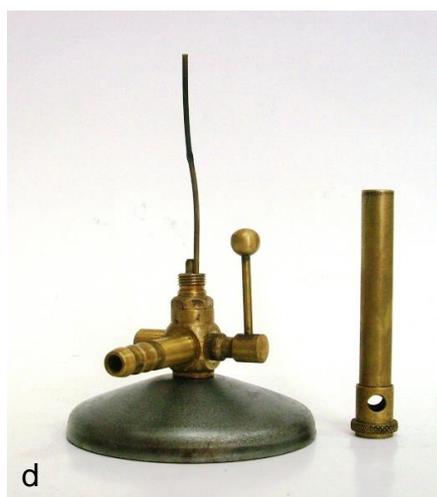

**Fig. 4.** a) Bunsen burner with a simple shape (6.5 x 11 x 9 cm); the collar for air regulation is missing. b) Bunsen burner with needle valve, for gas flow adjustment (8 x 8 x 12.5 cm); the needle valve is on the left and the hose-barb tube for the gas inlet is on the right. c) Bunsen burner with lever tap and auxiliary flame (8 x 8 x 12.5 cm); d) inner thin straw for auxiliary flame.

Figure 5 shows three Bunsen burners of the Museum of Chemistry; all of them are equipped with chimney to protect flame from wind blows, when used as heater. These burners probably date back to the end of XIX century. All the Bunsen burners have undergone only cleaning and conservation interventions to remove possible causes of deterioration.

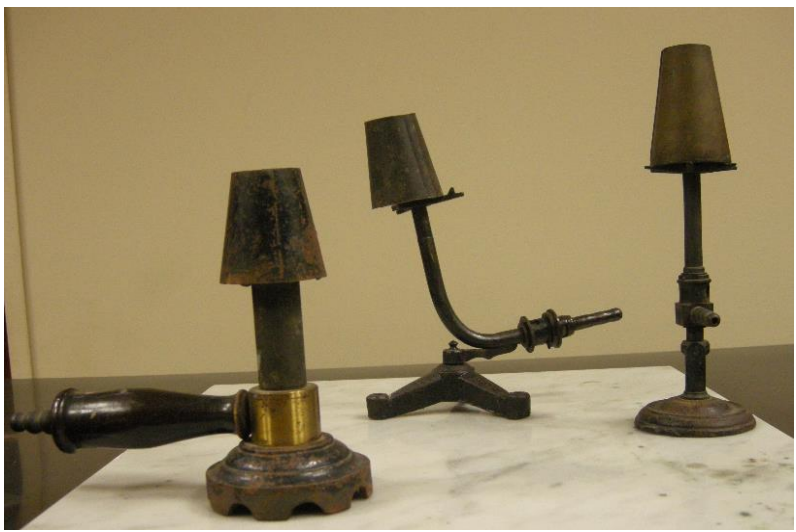

**Fig. 5.** Bunsen burners of the Museum of Chemistry.

## 3 Use of Bunsen burner for flame tests

Since its invention, the Bunsen burner has mainly been used for analytical purposes to perform experiments with spectroscopes, as shown in figure 6 (Kirchhoff & Bunsen, 1860). In chemistry laboratory, it is routinely used for a quick identification of chemical elements by the flame test technique. It consists in heating the substance by the burner and observing the resulting flame's color. The historical development of the analysis of chemical elements through the flame's colors is described in Jensen, 2014.

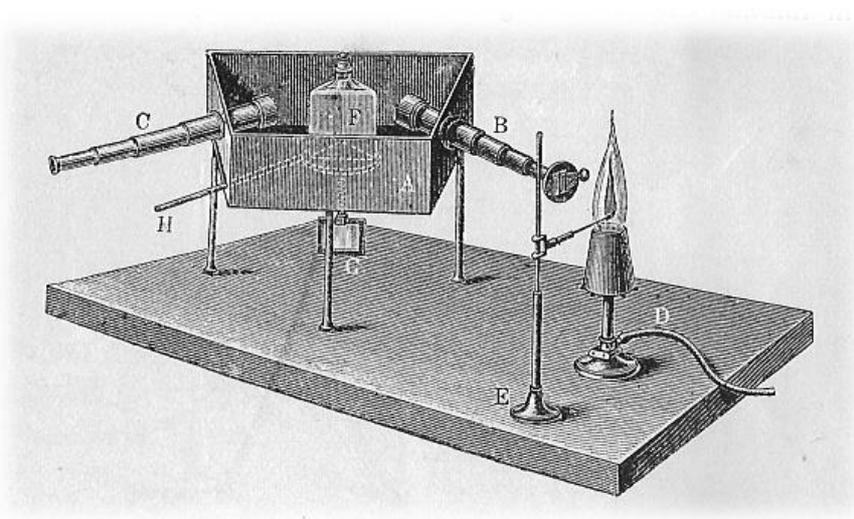

**Fig. 6.** Bunsen burner used with spectroscope, from Kirchhoff & Bunsen, 1860.

The experiment performed with students consisted in observing the different colors of the Bunsen burner flame produced by different chemical elements. A platinum wire was used to put a small amount of compound into the hottest part of the flame (a nickel-chrome wire can be used as well). The wire was previously cleaned, by immersing it into a solution of hydrochloric acid, HCl, and repeating the procedure until the flame became color less. Then, a dry sample was taken by the wire and introduced into the lower oxidizing zone of the flame (see fig. 3), which got the characteristic color of the element in the sample. Figure 7 shows the flame's colors given by the chlorides of the following elements: Sodium, Calcium, Potassium, Barium, Lithium, Strontium and Copper (from left to right).

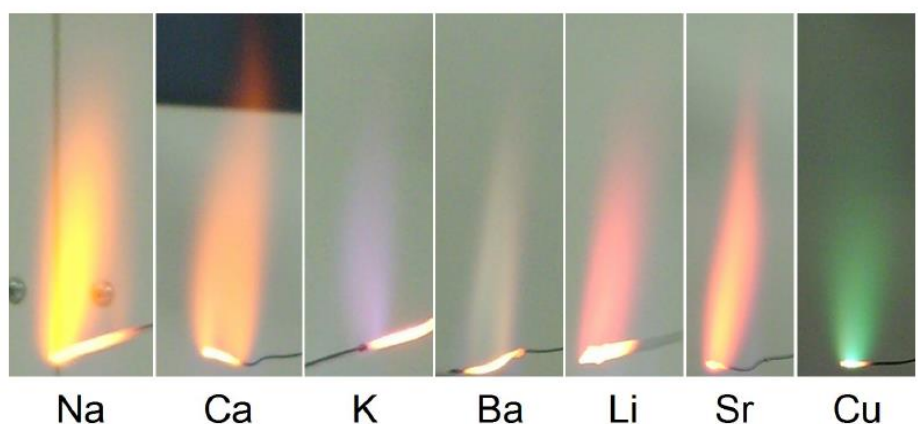

**Fig. 9.** Colored light emitted by the different elements when heated by the Bunsen burner flame. From left to right: Sodium (golden yellow), Calcium (brick-red), Potassium (lilac), Barium (yellowish green), Lithium (carmine red), Strontium (crimson red) and Copper (blue green).

Put in the hottest zone of the flame, solid salts volatilize and decompose into their chemical elements. The outer electrons in the quantum ground state $S_0$ are excited to a quantum state $S_1$ with higher energy. Subsequently, they return to the ground state emitting a photon of energy $h\nu$, where $h$ is the Planck constant and $\nu$ the frequency of the photon, due to electronic transitions between the different atomic energy levels $S_1$ and $S_0$. As a consequence, the non-luminous part of the flame gets the color that is determined by the electronic states of the elements contained in the substances put into the flame. To increase volatility, salts are dipped in concentrated hydrochloric acid, to be converted into the corresponding highly volatile chlorides, before being put into the flame. Furthermore, to enhance the diagnosis, the colored flame can be observed by a portable spectroscope, which allows one to observe the spectral composition of the emitted light. However, that is beyond the scope of the present paper.

## 4 Discussion

In the last decades, science museums played a more and more important role in science education, not only of students but also of the public at large. Indeed, early science museums were organized with rooms filled by glass cases containing instruments and apparatus. They largely served to house collections, as a resource for scholars. Conversely, last generation of science museums lost this characteristic, as well as the name "museum". Today science centers are endowed with intensive animation and hands-on exhibits that allow visitors to explore interactively science phenomena (Fridman, 2010).

The potential for educational approaches in science museums through the "Whole Science" approach and the "storytelling approach" are recently pursued, in which practical experiences are combined with theoretical considerations (Heering, 2017).

Although public science museums and science centers evolve toward a wider public education mission, University scientific museums maintain their original role to save and preserve the history and the cultural heritage related to the development of the specific discipline.

Besides the high educational role of science centers, we would emphasize the possibility to utilize laboratory activities, connected with historical scientific instruments of the University museums, for an inquiry-based multidisciplinary science education (Edwards, 1997). Since students are asked to answer many questions to understand how an instrument works and how it was used in the past, it helps them to construct knowledge from their own experiences with the instrument under study.

Because of the experimental work, students understand and learn the Physics and Chemistry arguments, connected with the activities carried out, in a much better way than from the usual didactics at school. They studied in order to better understand the Physics and Chemistry concepts at the root of the Bunsen burner and the experiments on flame tests.

## 5 Conclusion

We have described how University scientific museums can reach secondary-school students with a "learning-by-doing" approach to explain the physics and chemistry concepts, connected to the study, conservation and valorization of the Bunsen burners of the collections, utilizing the potentialities of museum activities, as keys to success, to give students more hands-on learning experiences. Furthermore, we have illustrated some interesting Bunsen burners of the Historical Collection of the Physics Instruments and the Museum of Chemistry at the University of Palermo.

## Acknowledgements

Work done under financial support of SiMuA, University of Palermo. The authors thank I. Sciacca for kindly organizing the experiment of flame tests, S. Licata and D. Capotummino for accompanying students.